\documentstyle[11pt,aas2pp4]{article}

\lefthead{Blakeslee and Tonry}
\righthead{Globular Clusters in Fornax}

\begin{document}

\title{Globular Clusters in Fornax: Does $M^0$ Depend on Environment?}

\author{John P. Blakeslee\altaffilmark{1} and John L. Tonry}
\affil{Department of Physics, 6-216, Massachusetts Institute of Technology,
77 Massachusetts Ave., Cambridge, MA 02139}
\authoremail{john@arneb.mit.edu}
\authoremail{jt@antares.mit.edu}

\altaffiltext{1}{Guest observer, Cerro Tololo Inter-American Observatory,
operated by AURA, Inc.\ under contract to the National Science Foundation.}

\begin{abstract}
We present the $V$-band globular cluster luminosity functions (GCLFs) of
the Fornax Cluster galaxies NGC~1344, NGC~1380, NGC~1399, and NGC~1404.
Our observations reach to $V = 24.8$, roughly one magnitude beyond the
GCLF turnover $m^0_V$, with $\sim90\%$ completeness.  From the
amplitude of the galaxy surface brightness fluctuations, we also estimate
the number of globular clusters fainter than this cutoff magnitude.
The GCLFs of these galaxies are well fitted
by Gaussians; the weighted means of their turnover magnitudes and
dispersions are $\langle m^0_V\rangle = 23.88 \pm 0.10$ mag and
$\langle\sigma\rangle = 1.35\pm 0.07$ mag.  The assumption of
a universal value for the absolute magnitude of the turnover
$M^0_V$ places the Fornax cluster
$0.13 \pm 0.11$~mag more distant than Virgo.  However, in light
of recent Cepheid and other high-precision distance measurements,
as well as ongoing HST observations
of GCLFs for the purpose of determining the extra-galactic distance scale,
we choose to re-examine the universal GCLF hypothesis.
Based on data from groups and clusters of galaxies, we
find evidence that $M^0_V$ becomes fainter as the local density
of galaxies increases.
We speculate on the possible cause of this trend;
if it is confirmed, GCLF observations will be less useful for determining
distances, but may provide important information for constraining theories
of star formation in primordial galaxy halos.

\end{abstract}


\vskip 0.65cm
\centerline{(Accepted for publication in
{\it The Astrophysical Journal Letters})~~}

\section{Introduction}

The globular cluster luminosity function\break (GCLF) is often
employed as a standard-candle distance indicator based on the assumption
a universal value for its mean, or turnover, magnitude $M^0$
(see \cite{j92} for a review of the method).
Until recently, it was impossible to apply the GCLF method to determine
the distances of galaxies further away than Virgo.  Now, with HST and
improvements in ground-based seeing and instrumentation, it
becomes potentially much more powerful for determining the extra-galactic
distance scale.  Furthermore, new Cepheid and other high-precision local
distance measures would allow for a firm calibration of the method.

Baum et al.\ (1995a,b) have used HST to observe the the globular clusters (GCs)
of the Coma galaxies NGC 4881 and IC~4051 down to $V=27.6$ and $V=28.4$,
respectively.  They derive values of the Hubble constant $H_0$
near 60 km/s/Mpc.
Also with HST, Whitmore et al. (1995) studied the GCs of the
extremely rich M87 system to two magnitudes beyond $M^0_V$
(the $V$-band GCLF turnover) and derived $H_0 = 78 \pm 11$ km/s/Mpc.
Only a small part of the discrepancy in derived $H_0$ values can be
accounted for by the different calibrations used by the two groups.
This situation leads
one to suspect that the GCLFs themselves may be intrinsically
different, especially as there remains no firmly established physical
basis for assuming a universal $M^0$.  Previously, there have been
suspicions that $M^0$ is different for spirals and ellipticals (\cite{sh93}),
with the root cause of this difference being metallicity variations
(Ashman, Conti, \& Zepf 1995), but $M^0_V$ was assumed not to vary among
large ellipticals.
As Whitmore et al.\ candidly remark, this ``crucial assumption'' of a
universal GCLF is ``a hypothesis that needs further verification.''

In this Letter, we examine the current state of the universal GCLF hypothesis.
First, we present new observations of GCs around four Fornax galaxies.
Fornax is an important cluster for testing distance determination methods, as
it is spatially much more concentrated than Virgo while being at nearly the
same distance (e.g. \cite{t91}; Ciardullo, Jacoby, \& Tonry 1993).
We find that the GCLF exhibits remarkably little variation
for these Fornax galaxies.
Next, we use independent distance measurements to galaxies and galaxy groups
to compare derived $M^0_V$ values in different environments.
We find somewhat startling 
evidence that $M^0_V$ becomes fainter as the local density of galaxies
increases. Further verification is once again needed, but if the observed
trend proves real, it would have major implications for the GCLF method of
distance measurement as well as for theories of GC formation.
We conclude with a discussion of these implications, in particular how
the local galaxy environment may govern the properties of GC populations.~~~~~

\section{Observations and Reductions}
We observed the Fornax Cluster galaxies NGC 1316, NGC 1344, NGC 1380,
NGC 1399, and NGC 1404 in 1995 August with the Tek $2048^2$~\#4 CCD detector
at the Cassegrain focus of the 4~m telescope at Cerro Tololo. Four 600~s
$V$-band exposures were taken of each galaxy, except NGC 1316, for which
five 600~s exposures were taken.  We also obtained 2400~s of integration
on a background field 1\fdg 5~west of the cD NGC 1399.
The image scale was 0\farcs158 pix$^{-1}$; however, the chip was
slightly vignetted around the edges, and we shifted the telescope
$\sim6^{\prime\prime}$ between individual exposures, so the final field size
which received the full integration time was about $5\farcm1 \times 5\farcm 1$.
We processed the images as described by Ajhar, Blakeslee, \& Tonry (1994)
and Blakeslee \& Tonry (1995; hereafter BT95).
The seeing in the final NGC 1399 image was 0\farcs94, while
the seeing in the other images ranged from 1\farcs03 to 1\farcs05.
The photometry was calibrated using Landolt (1992) standard stars;
there is no detectable Galactic
extinction in the directions of these galaxies (\cite{bh}).~~~

After subtracting smooth models of the galaxy surface brightness profiles,
we used a version of the program DoPHOT
(Schechter, Mateo, \& Saha 1993) for the point source photometry.
Completeness corrections were determined by scaling
and adding grids (so as to avoid artificial crowding)
of $32\times32$ ``cloned'' PSF stars and then
finding them again with DoPHOT.  The scaling was done at 0.4~mag intervals; we
interpolated to find the completeness corrections at intermediate magnitudes.
The images were then divided up into three radial regions:
20-40$^{\prime\prime}$, 40-80$^{\prime\prime}$, and 80-160$^{\prime\prime}$.
We settled on a cutoff magnitude $m_c$ of $V=24.4$ for the innermost
region in each galaxy; at this magnitude,
the completeness levels for this region ranged from 77\% to to 88\%.
For the intermediate region, the completeness levels were in the same range
at $V=24.8$, so we used this for the cutoff magnitude.  For simplicity, we also
used $V=24.8$ for $m_c$ in the outermost region, although the completeness
levels for this region ranged from 90\% to 96\% at this magnitude.
The photometric error is typically $\sim 0.12$~mag at $V=24.8$.
Any objects classified by DoPHOT as extended were excluded from further
analysis.  Objects brighter than $m_c$ in
each region were then binned in magnitude and our completeness
corrections were applied.  We subtracted the completeness-corrected
luminosity function of the unresolved objects in the background field
from the luminosity functions of the objects in the program fields
to produce the final GCLFs presented below.

After removing all objects brighter than $m_c$ in
each region, we measured the PSF-convolved variance, or fluctuations,
remaining in the image.  Our variance analysis method is described
in detail by Tonry et al.\ (1990).  The variance measurement effectively acts
as an ``extra bin'' in fitting the GCLF.
BT95 demonstrate how this measurement can be used for deriving
GCLF parameters.  The conversion from measured variances to GC densities is
done in the same way here, but the GC counts are treated differently
in that they are binned for more GCLF shape information.
In addition, we leave $m^0_V$ as a free parameter, instead of varying it only
within some presupposed acceptable range, as in BT95.
A more detailed account of our point source photometry and positions,
completeness experiments, and fluctuation measurements will be provided
elsewhere (\cite{betal}).

\font\tenrm=cmr10 
\font\fiverm=cmr5
\font\tinyrm=cmr8
\font\ninerm=cmr9
\font\verytinyrm=cmr9 at 6truept
\font\headrm=cmr10 at 7.3truept
\font\sevenrm=cmr7 at 7truept
\font\fiverm=cmr5 at 5truept
\font\eightrm=cmr8                              
\font\eightit=cmti8                             
\font\eightbf=cmbx8                             
\font\nineit=cmti9
\font\ninebf=cmbx9

\begin{figure}
\plotfiddle{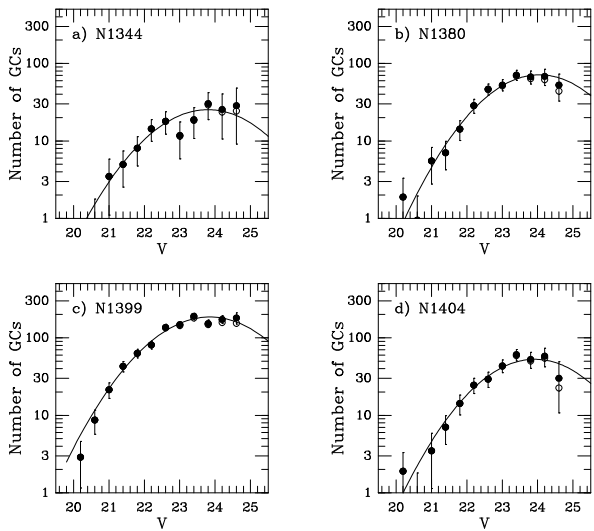}{2.4in}{270}{40}{40}{-113}{200}
\caption{
\tenrm\baselineskip 11pt
The globular cluster luminosity functions of the
four Fornax galaxies.  Filled symbols represent the final counts following
incompleteness and background corrections.  Open symbols show what
the counts would be with background subtraction, but no
incompleteness corrections.  Errorbars represent the uncertainties in our
corrections and the Poisson errors in the counts.
\label{fig1}}
\end{figure}

\begin{deluxetable}{lcrrrrrrc}
\small
\tablecaption{Final Fornax Gaussian GCLF parameters.\label{tbl1}}
\tablehead{
\colhead{Galaxy} & \colhead{Type}   & \colhead{$V_T$}   &
\colhead{$\sigma$}  & \colhead{$\pm$}  & \colhead{$m^0_V$} & \colhead{$\pm$} &
\colhead{\phm{1}N} & \colhead{$\pm$}
}
\startdata
N1344 & E5   &10.35 & 1.35 & 0.18 & 23.80 & 0.25 & 140 &20\phd \nl
N1380 & S0   & 9.91 & 1.30 & 0.17 & 24.05 & 0.25 & 375 &35\phd \nl
N1399 & E1/cD & 9.57 & 1.38 & 0.09 & 23.83 & 0.15 &1120 &45\phd \nl
N1404 & E1 & 9.98 & 1.32 & 0.14 & 23.92 & 0.20 & 300 &30\phd \nl
\enddata

\tablecomments{\tenrm\baselineskip 11pt
Columns list: galaxy name, Hubble type, total apparent $V$-band
magnitude (\cite{RC3}), Gaussian dispersion of the GCLF, $V$-band
GCLF turnover magnitude, and total number of GCs which went into the
Gaussian fits (following corrections for incompleteness and background).}
\end{deluxetable}

\section{Results}

We defer the analysis of the spatial structure and total sizes of the GC
populations and concentrate on the luminosity functions.
We also defer any further discussion of our observations of GCs in
the giant disturbed galaxy NGC~1316 (Fornax A).
The GCLF of this galaxy was not well fitted by a Gaussian model
($\chi^2 \gtrsim 3$ instead of $\sim 1$) and would require a more thorough
analysis than we can provide here.

\subsection{The GCLF in Fornax}

Figure~1 presents the $V$-band GCLFs of the four Fornax galaxies.
The plotted curves are Gaussians having $\sigma$ and $m^0_V$ values which
are the weighted means of the values found in our three analysis regions
of each galaxy.  The values were derived in the separate
regions by $\chi^2$ minimizations using the counts brighter
than the individual cutoff magnitudes and our variance measurements.
For purposes of displaying the GCLFs, however, we used $m_c=24.8$
everywhere, applied our completeness corrections, then binned the
regions all together.

Table~1 lists our final values for the GCLF parameters; they are
nearly identical within the errors. The cD NGC 1399 is the only one
with a well-studied GCLF.  Geisler \& Forte (1990) found
$m^0_V = 23.45$ for this galaxy, but assumed $\sigma = 1.20$~mag
(these parameters
are correlated when the limiting magnitude is near the turnover).
Bridges, Hanes, \& Harris (1991) used $\sigma \approx 1.40$, and found
$m^0_V = 23.85 \pm 0.30$, in close agreement with our value.

As weighted means of the Fornax GCLF parameters,
we take $\langle\sigma\rangle_{For} = 1.35\pm0.07$~mag;
$\langle m^0_V\rangle_{For} = 23.88\pm0.10$.
Whitmore et al. (1995) found $m^0_V = 23.72 \pm 0.06$ for M87.  We average
this with the values for NGC 4472 and NGC 4649 (\cite{sh93})
using $(B-V) = 0.75 \pm 0.05$ (\cite{cha90}, 1991) to find
$\langle m^0_V\rangle_{Vir} = 23.75\pm0.05$.  Thus, assuming a universal
$M^0_V$ yields $\Delta(m-M) = 0.13 \pm 0.11$ for the relative Fornax-Virgo
distance modulus.  If we were to include
NGC 4365 (\cite{sh93}) and NGC 4636 (\cite{kiss94}) in the Virgo average,
the relative modulus would drop by $\sim0.1$,
but both these galaxies are questionable Virgo members (\cite{nbg};
\cite{tal}).

\subsection{Does $M^0$ Depend on Environment?}

We now forsake the assumption of universality and compare the Fornax GCLF
with those observed elsewhere.
To do this, we need a self-consistent set of independent distance
determinations to galaxies or groups in which $m^0_V$ has been measured.
We start by fixing $(m-M) = 31.0$ as the Virgo distance modulus; this is both
the Jacoby et al.\ (1992) value and the latest HST Cepheid
Key Project result (\cite{f96}).  For the relative
Fornax-Virgo distance modulus, we take $\Delta(m-M) = 0.25 \pm 0.08$ from
an average of the PNLF, SBF, $D_n-\sigma$, SN~Ia, and Tully-Fisher methods
(\cite{cjt93}; \cite{7s89}; \cite{r96}; \cite{w96}).

We use measurements of $m^0_V$ for M31 (\cite{s92}), M81 (\cite{pr95}), the
Leo group ellipticals (\cite{h90}), and the HST limit on $m^0_V$ for
NGC 4881 in Coma (\cite{baum1}).
The Cepheid distance moduli to M31 and M81 are $24.43\pm0.10$ and
$27.80\pm0.20$, respectively (\cite{fm90}; \cite{m81hst}).
This M31 distance modulus is the proper one to use, as it is the one
assumed by Jacoby et al.\ (1992) and is consistent with the more recent
HST Cepheid distances.  For the Leo group, we average the HST
Cepheid distance to the spiral M96 (\cite{tan95}) with the PNLF and SBF
distances to the ellipticals (\cite{cjt93}) and get $(m-M) = 30.2 \pm 0.13$.
The relative Virgo-Coma distance modulus is well determined at
$\Delta(m-M) = 3.71 \pm 0.10$ (e.g. \cite{vdb92}; Whitmore et al.\ 1995).
Finally, in an effort to preserve neutrality in the controversy
over the RR Lyrae calibration (see \cite{vdb95}), we omit the
Milky Way from our discussion, remarking only that recent MW $M^0_V$ values
have ranged from $-7.29 \pm 0.13$ (\cite{s92}) to $-7.60 \pm 0.11$
(\cite{st95}; see also the discussion by \cite{baum1}).

Figure~2 shows the resulting $M^0_V$ values plotted against the the velocity
dispersions of the groups and clusters, from Tully (1987a) and Zabludoff,
Huchra, \& Geller (1990).  We use velocity dispersion as the most
convenient indicator of the depth of the local potential well; it closely
correlates with Tully's estimated group densities and with cluster
richness.  There is a trend of decreasing turnover
luminosity with increasing local density.  The offset in $M^0_V$ between
the small groups and Fornax/Virgo is 0.4~mag. The use of the
straight Cepheid distance to Leo would move its $M^0_V$ brighter by
0.1~mag, further away from the other ellipticals; the inclusion of
NGC 4365 and NGC 4636 would move the Virgo $M^0_V$ fainter.
In addition, the preliminary result $m_V^0 \approx 28.0$
($M_V^0 \approx -6.7$) for IC 4051 (Baum et al.\ 1995b)
indicates that there may be {\it another}\
$\sim0.5$~mag offset in $M^0_V$ for the very rich Coma cluster.
Thus, we believe we are seeing real evidence for an environmental
dependence of $M^0_V$.

\section{Discussion}

We have found that the GCLF is remarkably constant within
the Fornax cluster, but,
as Figure~2 shows, varies with environment.
Ashman et al.\ (1995) suggested that metallicity differences result in
$M^0_V$ values which are systematically brighter by $\sim0.15$~mag
for spirals.
Large ellipticals usually do have higher metallicity GC populations;
however, NGC 4881 in Coma has a GC color/metallicity distribution
similar to that of the MW (\cite{baum1}), yet its $M^0_V$ is very
faint.  In addition, the Leo elliptical
NGC 3379, with its relatively high metallicity GC population
(\cite{abt94}), has an exceedingly bright $M^0_V$, though with
a large uncertainty (\cite{h90}). Finally, we note that the
magnitude of the environmental effect we propose is
a factor of 3-6 larger than the Ashman et al. $M^0_V$ metallicity shift.

\begin{figure}
\epsscale{0.75}
\plotone{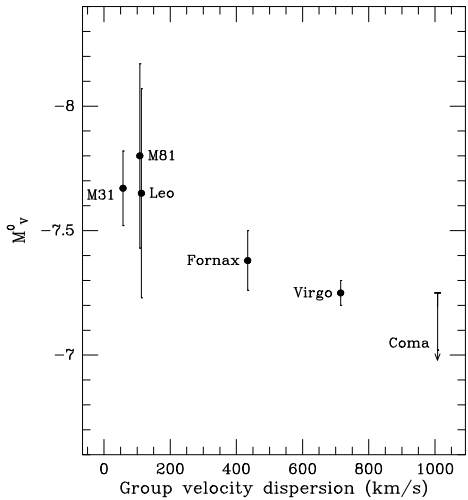}
\caption{\tenrm\baselineskip 11pt
The GCLF turnover magnitude $M^0_V$ plotted against
the velocity dispersion of the host galaxy's environment, used as a measure
of the local density.  See text for details.
\label{fig2}}
\end{figure}

\begin{figure}
\epsscale{0.85}
\plotone{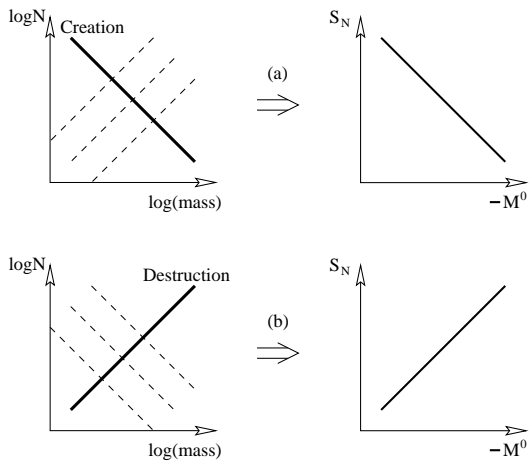}
\caption{
\tenrm\baselineskip 11pt
The effects of variable GC creation/de-struction mechanisms.
In part (a) the GC creation process, shown as a power-law
growing to smaller mass, is universal (dark solid line), and
the destruction process, a power law which wipes out low-mass objects,
varies with environment (dashed lines).  Since the total number of GCs
is the integral under the intersecting
creation/destruction lines, this situation results in
an anti-correlation between GC specific frequency $S_N$ and
GC mean logarithmic luminosity $-M^0$.
In part (b) the creation process (dashed lines) varies with environment,
while the destruction process (dark solid line) is universal.
The result here is a positive correlation between $S_N$ and $-M^0$.
\label{fig3}}
\end{figure}

We suggest that the most straightforward way to produce the
present-day near-Gaussian GCLF is to assume that two simultaneous and
competing effects were operating when GCs formed: a ``creation''
process which preferentially created low-mass GCs, cutting the
mass function off at the high end, and a ``destruction'' process which
inhibited the formation of, or quickly destroyed, low-mass GCs.
If each process operated in a manner which was independent of the
details of the environment, then the  final mass (luminosity) function
would be universal, but if one depended more sensitively on environment
than did the other, the final mass function would vary.
This situation is schematically illustrated in Figure~3, where we
use $S_N$ for the GC ``specific frequency''
(number of GCs per unit luminosity of the host galaxy).
In Figure~3a, we assume that the creation process is relatively
universal, but that the destruction process varies.  This leads to
$M^0$ being variable, and predicts an inverse correlation between
the number of GCs formed and their
mean brightness.  On the other hand, if the destruction process is
constant and the creation process is more variable we will again
get a variable $M^0$, but with a direct correlation between the
number of GCs and their luminosity, as shown in Figure~3b.

Empirically, we think we see evidence for the latter sort of behavior
among ``coeval'' galaxies, i.e., those located within the same
physical association.  In Virgo, for instance, M87 has a very
large $S_N$ and a slightly brighter $M^0$ than its close neighbors,
and similarly in Fornax for NGC~1399.
In this context, the GCLF would depend on the extent to which the host
galaxy dominated its local environment.
On the other hand, the main point of this paper is that we see the former
behavior among very ``heterogeneous'' systems of galaxies.  Young
groups dominated by spirals have fewer GCs than galaxy clusters such as Virgo,
which in turn may have fewer than rich clusters such as Coma, and we find
that the central luminosity of the GCLF is declining along this sequence.

As an example of how such an interplay of opposing processes might work
in practice, we consider the common picture of structure formation
through gravitational instability.
Here, the ``creation'' process is the primordial spectrum of density
fluctuations which favors low-mass clusters, and
the ``destruction'' process is the inhibition of the collapse
of low-mass objects resulting from the Jeans mass.
In this picture, the Jeans mass can be a very rapidly growing function
of time (Tegmark et al.\ 1996), and the densest systems of galaxies,
forming first, would have experienced a less restrictive low-mass cutoff
and hence have more, and fainter, GCs.
This is precisely the case depicted in Figure~3a.

Harris \& Pudritz (1994) have proposed a detailed astrophysical theory
of GC formation which is perhaps more illustrative of the ``coeval''
case of Figure~3b.  They suggest that the ``creation'' process is made
more efficient by the larger external pressures of dense environments.
Their ``destruction'' process is the tidal disruption and evaporation
of low-mass GCs, and this might be less sensitive to environment
(although see Murali \& Weinberg 1996).
Of course, if the cutoff is very abrupt (a steep
``destruction'' line), then $M^0$ will not correlate very
strongly with $S_N$.

We do not have better ``creation'' and ``destruction'' processes
to offer than have been advanced elsewhere, but we believe that this
description is a profitable way to frame the discussion. Until now, the
assumed constancy of $M^0$ has been a serious obstacle to reasonable
models for GC formation, so we conclude by re-emphasizing our primary point.
The GCLF apparently {\it does} depend on environment, with $M^0_V$ being
fainter in denser regions, although it may be
remarkably constant within a single group of galaxies.
This dependence will present challenges for
the use of the GCLF as a distance indicator.   On the
other hand, it opens the door for correlations between $M^0$ and
$S_N$, and $M^0$ and environment, which may yield valuable insights
into the conditions and processes which prevailed at the time of
GC/galaxy formation.

\acknowledgments

J.B. thanks Ed Ajhar for invaluable help while observing, as
well as profitable and sometimes overly animated discussions,
and the CTIO\break support staff for their tireless troubleshooting\break
efforts.  We also gratefully acknowledge helpful discussions
with Lam Hui and Paul Schechter.
This research was supported by NSF grant AST94-01519.


\end{document}